\documentclass{aa}  

\usepackage{natbib}
\usepackage{graphicx}
\usepackage{amsmath}
\usepackage{txfonts}
\usepackage{url}

\graphicspath{{./figures/}}

\begin{document}

\title{Cyclotron resonant scattering feature simulations}

\subtitle{I. Thermally averaged cyclotron scattering cross sections,
  mean free photon-path tables, and electron momentum sampling} 

\author{F.-W.~Schwarm\inst{1},
	G.~Sch\"onherr\inst{2},
	S.~Falkner\inst{1},
	K.~Pottschmidt\inst{3,4},
	M.~T.~Wolff\inst{5},
	P.~A.~Becker\inst{6},
	E.~Sokolova-Lapa\inst{9,10},
	D.~Klochkov\inst{11},
	C.~Ferrigno\inst{12},
	F.~F\"urst\inst{7},
	P.~B.~Hemphill\inst{8},
	D.~M.~Marcu-Cheatham\inst{3,4},
	T.~Dauser\inst{1},
	J.~Wilms\inst{1}
}
\authorrunning{F.-W. Schwarm et al.}

\institute{Dr. Karl Remeis-Sternwarte and Erlangen Centre for Astroparticle Physics, Sternwartstrasse 7, 96049 Bamberg, Germany
\and Leibniz-Institut f\"ur Astrophysik Potsdam (AIP), An der Sternwarte 16, 14482 Potsdam, Germany
\and CRESST, Department of Physics, and Center for Space Science and Technology, UMBC, Baltimore, MD 21250, USA
\and NASA Goddard Space Flight Center, Code 661, Greenbelt, MD 20771, USA
\and Space Science Division, Naval Research Laboratory, Washington, DC 20375-5352, USA
\and Department of Physics \& Astronomy, George Mason University, Fairfax, VA 22030-4444, USA
\and Cahill Center for Astronomy and Astrophysics, California Institute of Technology, Pasadena, CA 91125, USA
\and Center for Astrophysics and Space Sciences, University of California, San Diego, 9500 Gilman Dr., La Jolla, CA 92093-0424, USA
\and Faculty of Physics, M. V. Lomonosov Moscow State University, Leninskie Gory, Moscow 119991, Russia
\and Sternberg Astronomical Institute, Moscow M. V. Lomonosov State University, Universitetskij pr., 13, Moscow 119992, Russia
\and Institut f\"ur Astronomie und Astrophysik, Universit\"at T\"ubingen (IAAT), Sand 1, 72076 T\"ubingen, Germany
\and ISDC Data Center for Astrophysics, Universit\'e de Gen\`eve, chemin d'\'Ecogia 16, 1290 Versoix, Switzerland
}

\date{Received July 20, 2016; accepted August 30, 2016}

\abstract
{
	Electron cyclotron resonant scattering features (CRSFs) are observed
	as absorption-like lines in the spectra of X-ray pulsars.
        A significant fraction of the computing time for Monte Carlo
	simulations of these quantum mechanical features is spent on the
        calculation of the mean free path for each individual photon
        before scattering, since it involves a complex numerical
        integration over the scattering cross section and the
        (thermal) velocity distribution of the scattering electrons.
}
{
	We aim to numerically calculate interpolation tables which can be
        used in CRSF simulations to sample the mean free path of the
        scattering photon and the momentum of the scattering electron. The
	tables also contain all the information required for sampling
	the scattering electron's final spin.
}
{
	The tables were calculated using an adaptive Simpson
	integration scheme. The energy and angle grids were
	refined until a prescribed accuracy is reached.
	The tables are used by our simulation
	code to produce artificial CRSF spectra.
	The electron momenta sampled during these simulations were analyzed
	and justified using theoretically determined boundaries.
}
{
	We present a complete set of tables suited for mean free path
        calculations of Monte Carlo simulations of the cyclotron scattering
	process for conditions expected in typical X-ray pulsar accretion
	columns ($0.01 \le B/B_{\mathrm{crit}} \le 0.12$, where
        $B_\mathrm{crit}=4.413\times 10^{13}\,\mathrm{G}$, and 
	$3\,\mathrm{keV} \le k_\mathrm{B} T \le 15\,\mathrm{keV}$).
        The sampling of the tables is chosen such that the results
        have an estimated relative error of at most $1/15$ for all
        points in the grid. The tables are available online.
}
{}

\maketitle

\section{Introduction}\label{sec:intro}
Cyclotron resonant scattering features (CRSFs, often also called
``cyclotron lines'') have been measured in numerous accreting X-ray
pulsars and are the only direct way to measure a neutron star's
magnetic field. They result from the interaction of photons with
electrons in the presence of strong B-fields approaching the critical
field strength,
\begin{equation}\label{eq:B_crit}
B_\mathrm{crit}=
\frac{m_e^2 c^3}{e\hbar} = 4.413\times 10^{13}\,\mathrm{G}\,,
\end{equation}
where $m_e$ is the electron rest mass, $e$ its charge, and $c$ the
speed of light. Their positions can be estimated as
$\sim$$n\,E_\mathrm{cyc}\;(n=1, 2, 3,\ldots)$ using the 12-B-12 rule,
$E_\mathrm{cyc} \approx 12 B_{12}\,\mathrm{keV}$, with the B-field
strength $B_{12}$ given in units of $10^{12}\,G$. Cyclotron lines have
their origin in transitions of electrons between different Landau
levels, which are the discrete energy states an electron can occupy
within such a strong magnetic field.

The electrons are quantized perpendicular to the field and therefore
give rise to quantum mechanical absorption and resonant scattering
processes altering the spectral and spatial distribution of the
participating photons. The probability for an interaction to occur is
given by the corresponding cyclotron cross section. In the course of
simulating this process with Monte Carlo (MC) methods, these cross
sections can be used to sample the mean free path of a photon within
such a medium. We have separated this core issue from the larger
simulation code to allow for an efficient simulation of any complex
X-ray pulsar geometry based on precalculated tables of the mean free
path.

In the following we describe the calculation method and usage of these
mean free path interpolation tables and discuss how the sampling of
electron parallel momenta influences the formation of CRSFs. In
Sect.~\ref{sec:tables} we discuss the necessity of mean free path
interpolation tables and their usage and introduce their computation
and the interpolation mechanism. In Sect.~\ref{sec:momentum} we
explain the importance of the sampling of the electron parallel
momentum. In particular we illustrate the connection between cyclotron
resonances and the corresponding behavior of the sampled electron
parallel momenta, since the understanding of these parameters is
essential for the application of the Monte Carlo code to generate
synthetic spectra. Many more applications can be envisioned, including
the simulation of the influence of cyclotron scattering on the
electrons, or the overall accretion geometry. Here, we restrict
ourselves to the discussion of the mean free path interpolation
tables. Their motivation and application is described against the
background of Monte Carlo simulation of cyclotron lines. The
description and application of the full MC scattering code, which has
been written with the prime goal of imprinting cyclotron lines on the
continuum emission of astronomical X-ray sources, and which includes a
working fit model, will be the subject of a forthcoming publication
\citep[hereafter paper II]{schwarm16b}. Compared to previous MC
simulations, it allows for much more complex physical scenarios, the
exploration of which is the goal of this series of papers.

\section{Mean free path interpolation tables}\label{sec:tables}
\subsection{Motivation}
When simulating synthetic cyclotron line spectra using an MC method, a
photon with an initial energy is generated. Then the optical depth
$\tau$ to be traveled by the photon is drawn from the exponential
distribution $\exp(-\tau / \lambda)$, and converted into a geometric
path length. This requires the calculation of the mean free path
$\lambda$, that is, the inverse thermally averaged scattering cross
section. The photon is then propagated over this distance and the
scattering process is simulated. This simulation requires us to choose
an electron that has properties appropriate for this mean free path
(MFP). Paper II and \citet{schwarm12} show a flow chart of the full MC
scattering process. Since the calculation of the geometric path and
the scattering simulation are very time consuming, we have developed a
tabular interpolation scheme for the mean free path and electron
parallel momentum sampling to save computing time. It works on
precalculated tables, which were obtained using an adaptive process
refining the table until the interpolation error is smaller than a
preset limit.

The mean free path of a photon in a CRSF medium is the inverse of the
sum over the cross sections of all possible CRSF related interactions
between the photon and its possible scattering partners, which
throughout this work are assumed to be only electrons (see
Eq.~\eqref{eq:lambda} below). This calculation not only involves a
summation over all possible final Landau levels and spin states of the
electrons, but also a summation over all possible intermediate states.
Furthermore, the electrons have a temperature dependent continuous
momentum distribution parallel to the magnetic field, which leads to
Lorentz boosting of the scattering photons in the electrons' rest
frames giving rise to an integration over possible initial electron
parallel momenta as well. Finally, the cross sections are summed over
final polarization states and averaged over the initial ones for a
polarization averaged mean free path. Photons are either in ordinary
or extra-ordinary polarization mode. The modes differ in the
orientation of a photon's electric field vector with respect to the
plane defined by the direction of motion and the external magnetic
field \citep{canuto71,meszaros78a,becker07}.

Calculating the mean free path is only the first step in the MC
sampling process, but replacing it by an interpolation scheme has a
very large impact on the overall simulation time. The CRSFs can form
at rather small optical depths due to the resonant nature of the cross
sections. The small optical depths as considered here mean that most
of the injected seed photons escape the CRSF medium immediately. Only
1--10\% of the initial photons will interact with the medium via
cyclotron scattering. Therefore the only evaluation necessary for the
majority of photons is the one for their initial mean free path.

The simulations show that interacting photons tend to scatter around
the resonances in energy space until an electron with parallel
momentum that deviates sufficiently from the resonance condition is
hit. Spawned photons, that is, photons emitted by previously excited
electrons during their transition to a lower Landau level, are also
generated close to the resonances. This leads to a large number of
resonant photons compared to continuum photons and further motivates
the usage of the interpolation tables presented here since the
calculation time tends to increase significantly near the resonances.

For each interaction between photons and electrons, the parallel
momentum of the electron must be sampled according to the
corresponding thermal momentum distribution. For a given photon energy
and angle, the electrons with momenta that cause the photon to be
resonant in the electron rest frame have a much higher scattering
probability. Therefore the cyclotron resonances manifest in parallel
momentum resonances of the scattering electrons (see
Sect.~\ref{sec:momentum}).

The mean free path tables presented here store the total mean free
paths as well as the probability distributions needed for sampling the
electron momentum. Interpolating from these tables speeds up our
simulation by a factor of $\sim$60 compared to the case where all data
needed during the simulation is calculated without resorting to
interpolation tables. The difference between interpolation and
calculation becomes even larger, if the tables are used outside of the
simulation to simply interpolate mean free paths for several energies
without performing any other simulation steps. In this case
interpolation is faster by a factor of $\sim$2400 for input angles
almost perpendicular to the magnetic field.

\subsection{Calculation of the averaged cross sections}

The calculation of the mean free path needed for the sampling of
propagation lengths of photons within a scattering medium relies on
the integration of the scattering cross section $\sigma$ over a range
of electron momenta $p$, effectively averaging the cross section over
the parallel electron temperature \citep{daugherty86},
\begin{equation}\label{eq:profile}
\langle\sigma(\omega,\mu)\rangle_{f_\mathrm{e}}=\int_{-\infty}^{+\infty}\mathrm{d}p\,
f_\mathrm{e}(p,T)(1-\mu\beta)\sigma_\mathrm{rf}(\omega_\mathrm{rf},\mu_\mathrm{rf})\,,
\end{equation}
where $\mu = \cos\vartheta$, $\vartheta$ is the angle between the
photon's direction and the magnetic field, $\omega$ is the photon
energy, and $\beta = v/c$. The subscript ``rf'' refers to values in
the rest frame of the electron. $f_\mathrm{e}(p;T)$ is the electron's
momentum distribution, chosen here as a relativistic Maxwellian with
temperature $T$:
\begin{equation}
  \label{eq:f_p}
f_\mathrm{e}(p,T) \propto \exp\left( -\frac{1}{k_\mathrm{B} T} \left(\sqrt{m_e^2c^4+p^2c^2}-m_ec^2\right) \right)\,.
\end{equation}

The mean free path, $\lambda$, is then given by \citep{araya99}
\begin{equation}\label{eq:lambda}
\lambda(\omega,\mu) = 1/\langle\sigma(\omega,\mu)\rangle_{f_{\mathrm{e}}}\,.
\end{equation}

\begin{figure}\centering
  \resizebox{\hsize}{!}{\includegraphics{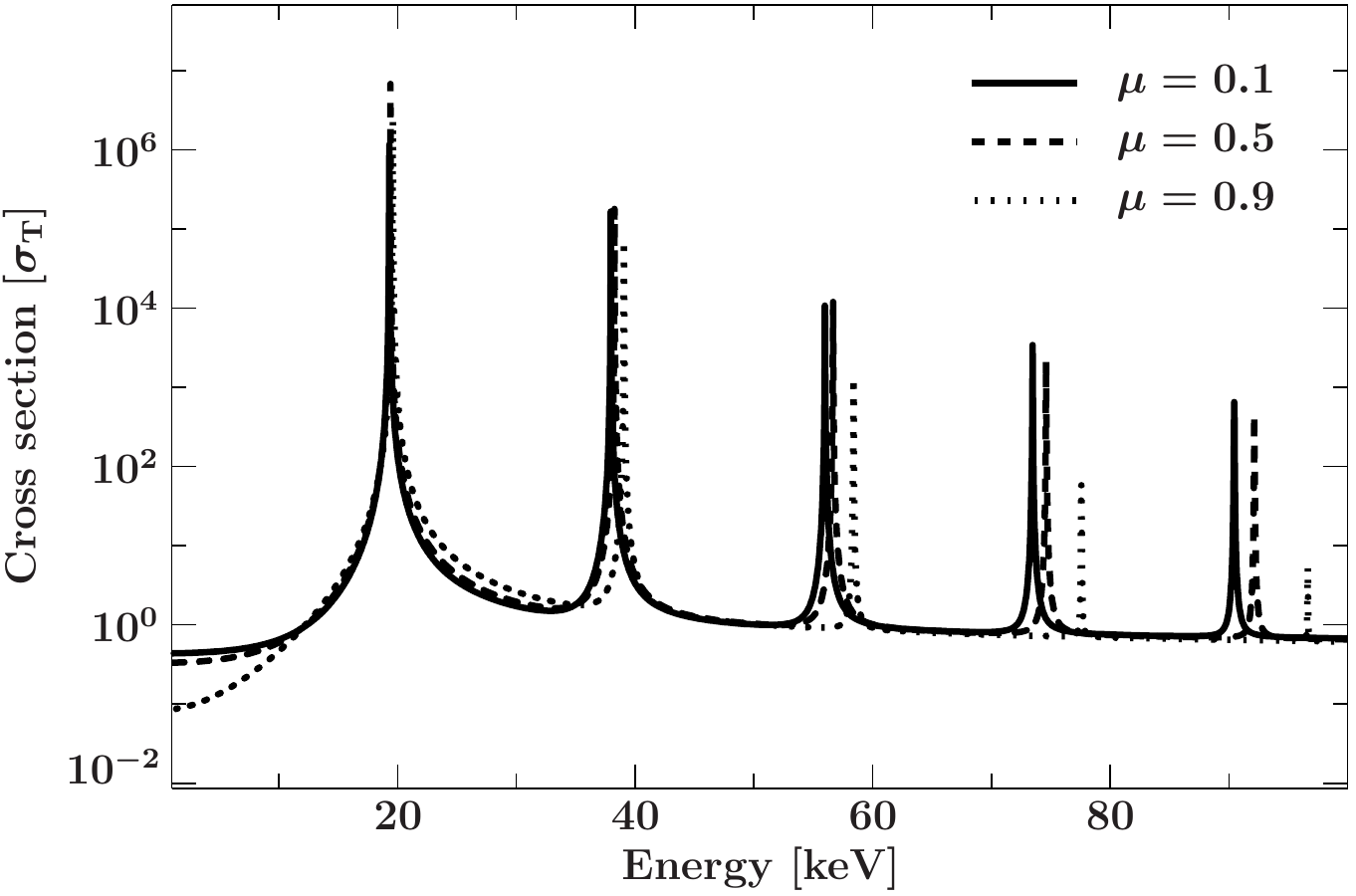}}
  \caption{Scattering cross section as a function of rest frame energy
    for exciting an electron in initial Landau level $n_i=0$ with spin $-1/2$ to
    level $n_f$ with $n_i \leq n_f \leq 6$ and final spin $-1/2$ within
    a magnetic field of strength $B = 0.0385\ B_{\mathrm{crit}}$,
    resulting in a fundamental cyclotron resonance at $\sim$20$\,\mathrm{keV}$.
    The solid, dashed, and dotted lines show cross sections for
    $\mu = \cos\vartheta = 0.1, 0.5, 0.9$,
    corresponding to angles between the path of the incoming photon
    and the $B$-field of $84\fdg2$, $60^\circ$, and
    $26\fdg8$, respectively. }\label{fig:scs}
\end{figure}

The fully relativistic differential Compton scattering cross section,
$\sigma_\mathrm{rf}$, used here was derived by \citet{bussard86} and
\citet{sina96}. These authors used the Breit-Wigner broadening
approximation and electron wave functions from \citet{sokolov68},
which are consistent with the perturbation theoretic order of the
calculation and the expected ``time dilation'' relation
\citep{graziani93}. \citet{daugherty78} presented an alternative
derivation of the cross sections using electron wave functions from
\citet{johnson49}. See \citet{herold82} for the relativistic
transition rates, which are needed for the cross section calculation
and for the sampling of the final Landau level into which an initially
excited electron decays.

We use a revised implementation of the code developed by
\citet{sina96} for the calculation of the scattering cross sections.
His original code is used in the MC simulation from \citet{araya97},
which has been employed by \citet{schoenherr07} to calculate the
Green's functions necessary for their CRSF fitting model. The
thermally averaged cross sections used in the \citet{araya97} code
differ from the ones calculated by \citet{harding91}
\citep{schwarm12}. We verified that the origin of this difference lays
in the integration of the cross section code into the MC simulation
rather than the cross section calculation itself, by succeeding to
reproduce the \citet{harding91} profiles with our revised
implementation of the \citet{sina96} code. The code has been
restructured to minimize code repetition and increase readability. It
has been generalized to allow for previously hard-coded variables,
such as numerical integration boundaries and methods, or the maximum
number of Landau levels taken into account, to be changed dynamically.
It has been extended to arbitrary temperatures by including an
adaptive Simpson integration scheme for averaging the cross sections
over the thermal electron momentum distribution. Improved error
handling and the addition of warning messages result in a more robust
implementation as needed for the time consuming MFP table
calculations. Unit tests ensure that the resulting cross sections are
in agreement with the ones calculated by the original code. The
resulting cross sections are also in agreement with the work of
\citet{gonthier14}. These authors derived the cross section for the
special, $\theta = 0$, case of photons propagating parallel to the
magnetic field, following the approach of \citet{sina96} as well.

The cross sections are sharply peaked functions, which are difficult
to integrate numerically. Figure~\ref{fig:scs} shows an example for
transitions from the Landau ground state with initial spin $-1/2$ to a
final state with negative spin summed over the first five possible
final Landau levels. A slight shift of the position of the resonances
to higher energies occurs for smaller angles of the photon to the
magnetic field.

The numerical evaluation of Eq.~\eqref{eq:profile} is closely related
to the evaluation of the probability distribution of the electron
momentum \citep{araya99},
\begin{equation}
  \label{eq:F_p}
F(p)\propto\int_{-\infty}^{p}\mathrm{d}p'\,
f_\mathrm{e}(p',T)(1-\mu\beta)\sigma_\mathrm{rf}(\omega_\mathrm{rf},\mu_\mathrm{rf})\,,
\end{equation}
which is required in Monte Carlo simulations in order to find the
momentum of the photon's scattering partner. The total inverse mean
free path, calculated by Eq.~\eqref{eq:profile}, accounts for all
possible electron momenta and is used to normalize $F(p)$ to unity. By
searching for the momentum $p$ for which $F(p) = \mathrm{Rn}$, with a
random number $0 \le \mathrm{Rn} \le 1$, the parallel momentum of the
scattering electron can be sampled.

In order to calculate $F(p)$ for a given magnetic field, temperature,
photon energy, and angle, we use an adaptive Simpson integrator
\citep{mckeeman62}. The thermal momentum distribution
$f_\mathrm{e}(p,T)$ becomes very small for large absolute values of
the electron momentum. Therefore the numerical integration limits were
set to $\sim$$\pm m_e c^2$. They cover a range much larger than the
expected momentum range of the electrons. This way the $45
k_\mathrm{B}T$ boundaries used by, for example, \citet{araya99} are
also covered for the temperature range of $3$\,keV to $15$\,keV. In
our adaptive approach the integration interval is successively split
into smaller intervals. The integrals of these intervals are
approximated by integrating a suitable third order polynomial. Due to
the adaptive nature of the integration method, the larger momentum
range we use compared to earlier works leads to only marginally
increased computing time.

The splitting of the integration intervals has to be stopped when a
suitable convergence criterion is fulfilled. \citet{lyness69} shows
that in fifth order approximation the maximum error of the integrator
can be estimated as
\begin{equation}\label{eq:simpson_convcrit}
  \epsilon = \frac{1}{15} \sum_i F_{a_i}^{c_i} - (F_{a_i}^{b_i} + F_{b_i}^{c_i})\,,
\end{equation}
where $F_{a_i}^{c_i}$ and $(F_{a_i}^{b_i} + F_{b_i}^{c_i})$ are the
numerical estimates for the integrals over the interval $[a_i, c_i]$
before and after bisecting the interval. We choose a relative maximum
error of $1/15$ as a convenient compromise between accuracy,
calculation time, and table size. The error is halved in each
recursion step to ensure that the total error estimate remains below
the chosen maximum error for the total interval.

\subsection{Optimizing table size: adaptive calculation}
\begin{figure}\centering
  \resizebox{\hsize}{!}{\includegraphics{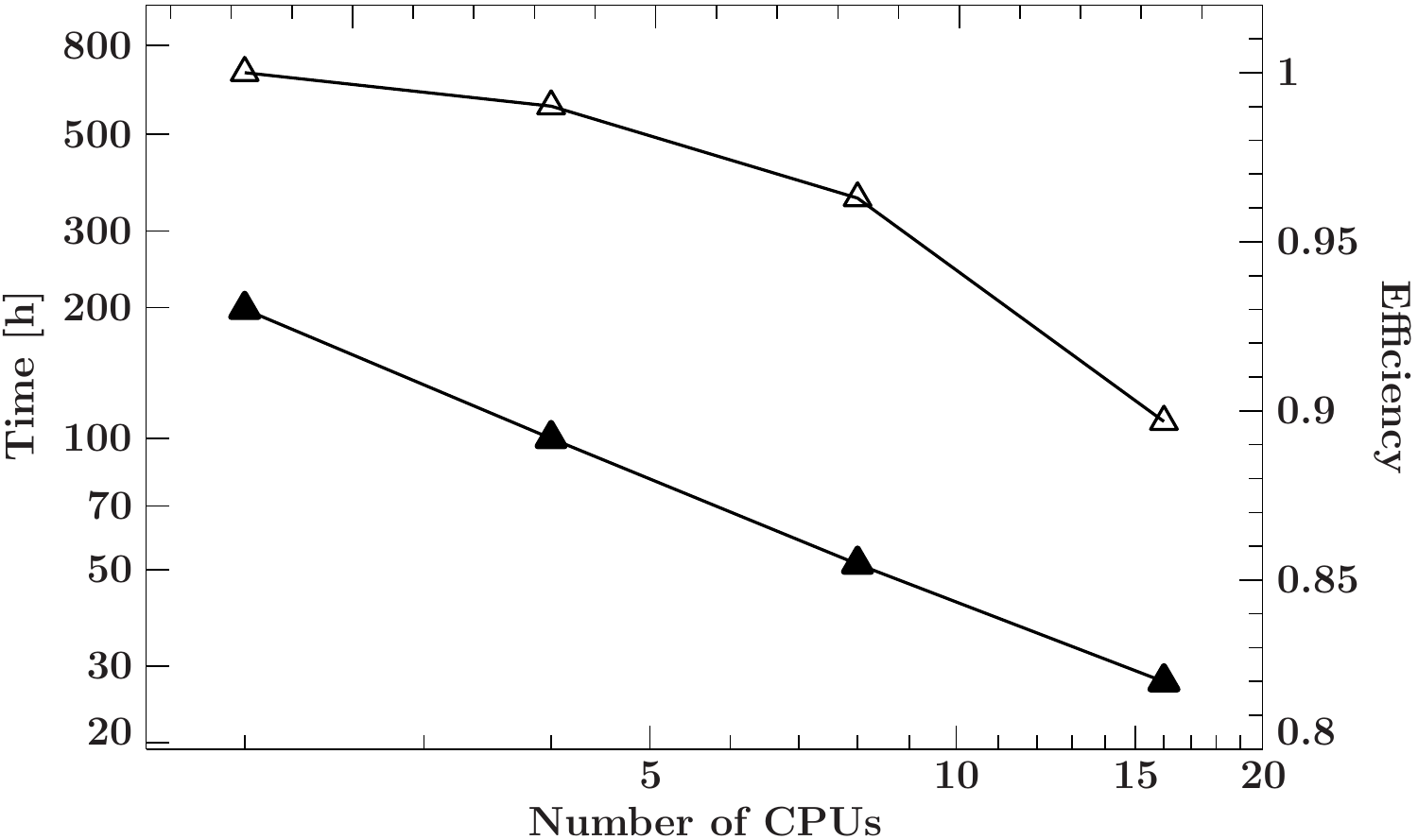}}
  \caption{Absolute CPU time (filled symbols) on the left axis and
    multi-core efficiency (open symbols) on the right axis for MFP
    table calculation. The efficiency is calculated as $\eta = t_1 /
    (n t_n)$, where $t_{n}$ is the execution time on $n$ CPUs. $t_1 =
    2 t_2$ is assumed since performing the calculation on only one
    core is too time consuming. The MFP table calculations were
    performed on an AMD Opteron 2.2\,GHz system. The execution times
    are for a magnetic field $B = 0.12 B_\mathrm{crit}$ and a
    temperature $k_\mathrm{B}T = 3$\,keV with a maximum error of 2/15.
    The CPU time scales with the table file size (see
    Sect.~\ref{sec:mfp_info}). }\label{fig:mpi_speed}
\end{figure}

The adaptive nature of this integration method takes advantage of the
fact that only few integration points are needed outside the
resonances. These choices allow for a computationally manageable
creation of mean free path tables on a time scale of 3000--30000\,CPU
hours on a typical work station with processor speed of 2--3\,GHz,
depending on the magnetic field, the temperature, and the error
tolerance. These calculation times are achieved by utilizing a few
dozen CPU cores\footnote{For the sake of simplicity the terms ``CPU''
  and ``CPU core'' are used equivalently throughout this work.} at the
same time using the Message Passing Interface \citep[MPI;][]{mpi94}.
Parallelization leads to a significant speed up by a factor of up to
$0.8 N_\mathrm{CPU}$ with respect to earlier implementations, where
$N_\mathrm{CPU}$ is the number of CPUs used.
Figure~\ref{fig:mpi_speed} shows this speed up as a function of
$N_\mathrm{CPU}$ in terms of the parallel computing efficiency, that
is, the total CPU time used for a mean free path table calculation or
for a simulation divided by the CPU time needed on only one CPU.

\begin{figure}
  \resizebox{\hsize}{!}{\includegraphics{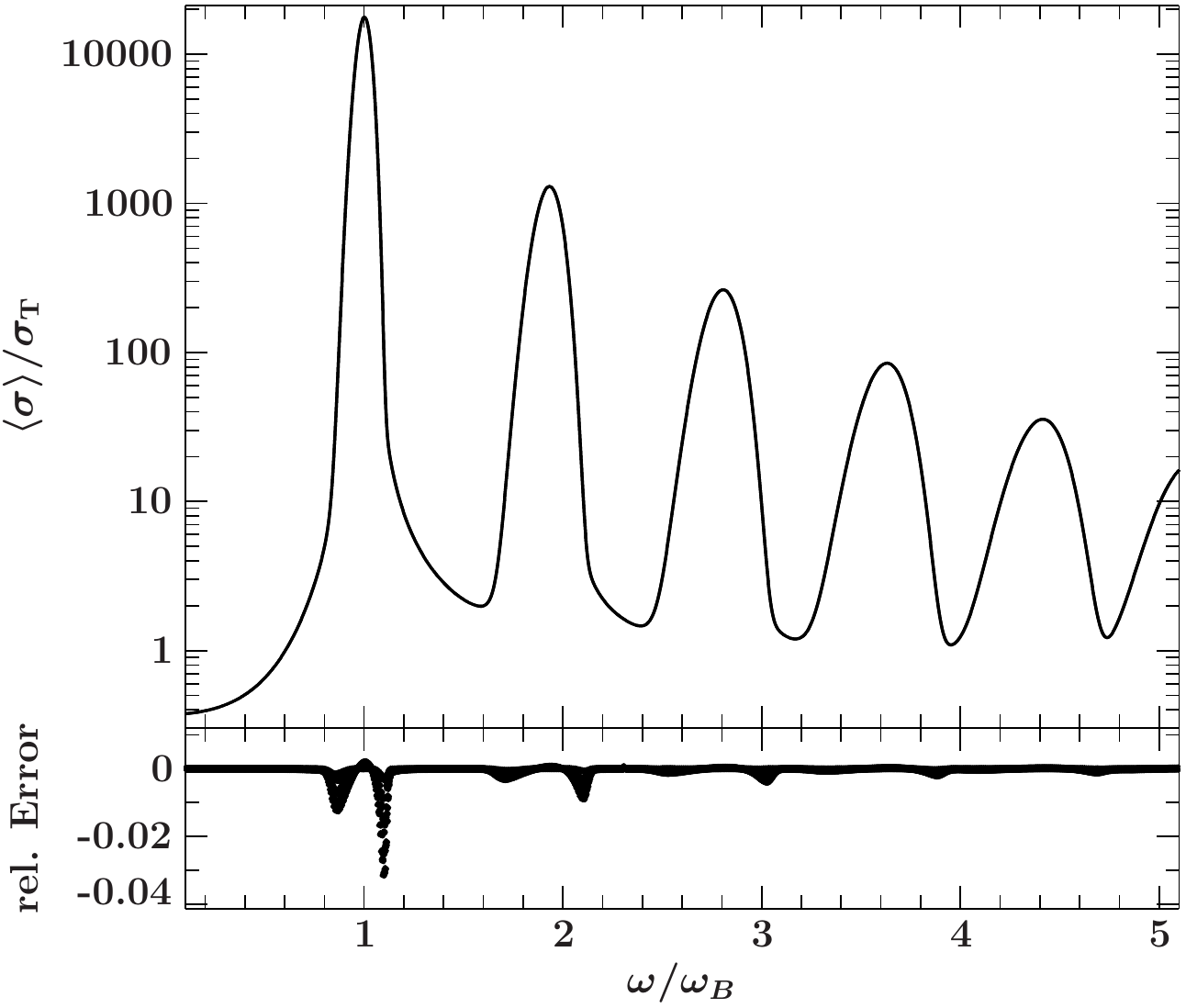}}
  \caption{Thermally averaged cyclotron scattering cross sections in
    units of the Thomson cross section $\sigma_\mathrm{T}$
    interpolated for $B=0.12\,B_\mathrm{crit}$,
    $k_\mathrm{B}T=3$\,keV, and angle $\vartheta=60^\circ$. The
    relative deviations to the calculated profiles, shown in the lower
    panel, are not exceeding the maximum error of 1/15 of the MFP
    table used for the interpolation.}\label{fig:mfp_errors}
\end{figure}

As discussed in Appendix~\ref{sec:interpo}, for the fast interpolation
of the mean free path we evaluate the integrals using another adaptive
process where the energy and the angular grid points are iteratively
refined until the mean free paths interpolated from the table do not
deviate from the corresponding calculated ones by more than a given
maximum deviation. After each calculation step of a new energy grid
point, the result is compared to the value obtained by linear
interpolation. The relative error of the interpolation is determined
and used as a convergence criterion. A minimum recursion depth as well
as maximal angle and energy differences prevent premature convergence.

The partial integrals, forming the cumulative distribution function
(CDF) of the electron momentum, are stored in FITS binary tables
allowing for efficient loading into a simulation. FITS tables can be
opened and read by many processes in parallel, enabling the usage of
parallel computation in simulations. Another advantage of the FITS
format is the efficient caching provided by modern FITS libraries such
as
\texttt{cfitsio}\footnote{\url{http://heasarc.nasa.gov/fitsio/fitsio.html}},
which minimize the bottleneck of disk reading operations. Although the
whole size of such an interpolation table is on the order of
1--200\,GB for the currently required accuracy, and thus will not
completely fit into memory for most common computers, for a given
setup the resonant nature of the scattering process means that the
most frequently required mean free paths will be from a small fraction
of the whole table only, which fits completely into memory and
therefore reduces the number of disk reads dramatically.

For testing the accuracy of the interpolation from the resulting
tables, we calculated the thermally averaged scattering cross section
profiles (Eq.~\ref{eq:profile}) on a much finer grid than the
tabulated one ($\sim$700--1800 energy grid points).
Figure~\ref{fig:mfp_errors} visualizes the profile for a magnetic
field of $B = 0.12 \, B_\mathrm{crit}$, interpolated on a grid of 5000
energy grid points. The relative deviations from the calculated values
do not exceed the relative maximum error of $1/15$.

Appendix~\ref{Appendix:mfptables} shows interpolated profiles for the
parameter combinations for which tables are provided. They are plotted
as a function of frequency and scattering angle for several $B$-fields
and temperatures in Figures~\ref{fig:mfp_tables}
and~\ref{fig:mfp_profiles}.

\subsection{Parameter ranges and file sizes}\label{sec:mfp_info}
The electronic data provided with this publication is available
online\footnote{\url{http://www.sternwarte.uni-erlangen.de/research/cyclo}}.
We provide tables for the mean free path calculated for $B$-fields and
temperatures covering the parameter ranges typical for accreting X-ray
pulsars, namely $B=0.01$, 0.03, 0.06, 0.09, and
$0.12\ B_\mathrm{crit}$ and $k_\mathrm{B}T = 3$, 6, 9, 12, and
15\,keV. Table~\ref{table:mfp_list} shows a list of the available
tables and their uncompressed file size. As shown by \citet{araya99},
the low continuum optical depth and the small collisional excitation
rate \citep{bonazzola79,langer81} of a typical accretion column,
combined with the high radiative cyclotron de-excitation rate
\citep{latal86}, imply that we can assume that all electrons are
initially in the ground state ($n=1$) Landau level and have a
relativistic thermal momentum distribution, which is not altered by
cyclotron resonant scattering. We therefore provide tables for this
case only. The full data set has an uncompressed size of approximately
2.7\,Terabyte. Compressing the tables using \texttt{gzip} leads to a
size reduced by roughly 50\%.

\begin{figure}\centering
  \resizebox{\hsize}{!}{\includegraphics{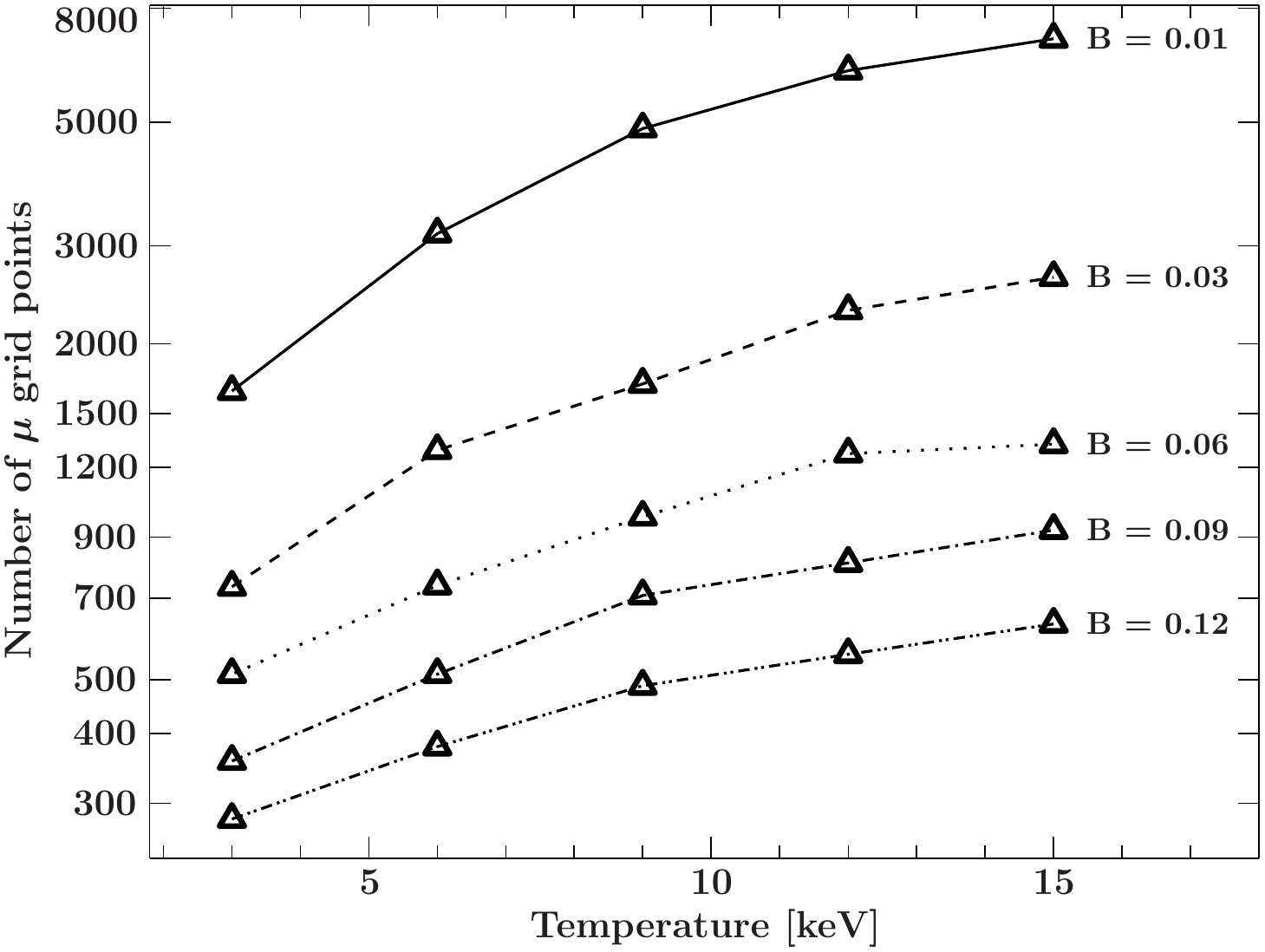}}
  \caption{ Number of $\mu$-grid points with respect to the electron
    temperature. Each line corresponds to one magnetic field strength.
  }\label{fig:mfp_info}
\end{figure}

The number of momentum grid points needed for convergence of the
numerical integration of Eq.~\eqref{eq:profile}, correlates positively
with the resulting inverse mean free path. Thus broader profiles and
those with larger absolute values require more $p$-grid points. The
number of energy grid points needed for accurate interpolation
strongly depends on the complexity of the profile and therefore on the
scattering angle. The sharply peaked profiles for angles
$\sim$$90^\circ$ to the magnetic field require many more angle grid
points. The average number of energy grid points for each table does
not depend on the temperature, nor does it vary with the magnetic
field strength for all but the highest field, $B =
0.12\,B_\mathrm{crit}$. For this field the tables have twice as many
energy grid points as the other tables. The number of angle grid
points in each table is correlated with the temperature and the
magnetic field strength. More $\mu$-grid points are needed for higher
temperatures since the broadening of the profiles towards larger $\mu$
increases with temperature. The negative correlation with magnetic
field strength can be explained by the larger absolute values of the
corresponding profiles. Figure~\ref{fig:mfp_info} shows the number of
$\mu$-grid points, in the set of MFP tables presented here, as a
function of the electron temperature and the magnetic field strength.
The largest table was obtained for the lowest magnetic field strength
and the highest temperature, and is 238\,GB in size.

\section{The sampling of the electron parallel momentum}\label{sec:momentum}

\begin{figure}\centering
  \resizebox{\hsize}{!}{\includegraphics{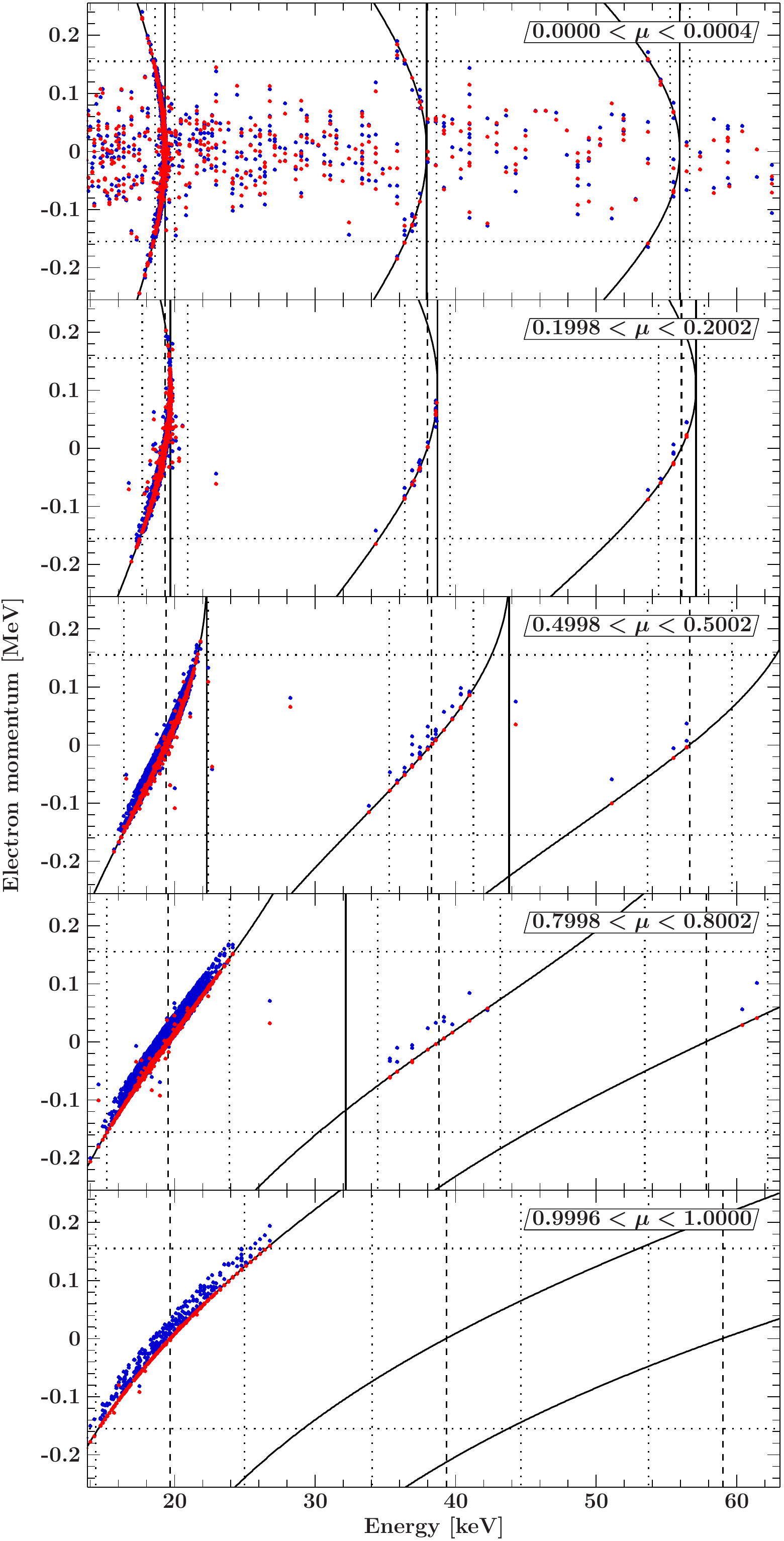}}
  \caption{ Sampled electron parallel momenta for one MC run. Each
    point represents one scattered photon by its energy on the
    $x$-axis and the corresponding electron's momentum before (red)
    and after (blue) the scattering process on the $y$-axis.
    Scattering events involving photons incoming at angles of
    $\sim$$90^\circ$, $79^\circ$, $60^\circ$, $37^\circ$, and
    $1^\circ$ to the magnetic field (from top to bottom) are shown.
    The setup of the calculation is the same as in \citet{isenberg98}
    with a magnetic field $B=0.0385\,B_\mathrm{crit}$, electron
    parallel temperature $k_\mathrm{B}T=5\,\mathrm{keV}$, and a
    bottom-illuminated slab geometry of the line-forming region.
    Vertical dashed lines show the cyclotron line energy, which is
    slightly shifted to higher energies for smaller angles to the
    magnetic field, because of the angular dependence of the resonance
    energy. This shift is often neglected by observers in favor of
    using the simplified 12-B-12 rule. Solid vertical lines mark the
    position of the cut-off energy beyond which no resonant scattering
    is possible. Dotted horizontal lines enclose the electron momentum
    range in which 99\% of the electrons are located (assuming a
    relativistic Maxwellian, Eq.~\eqref{eq:f_p}). Solid curved lines
    show the possible solutions for resonant scattering. The vertical
    dotted lines indicate the Doppler width of the cyclotron lines
    (see Eq.~\eqref{eq:E_width}). }\label{fig:pi}
\end{figure}

The electrons' momenta perpendicular to the magnetic field are
quantized. The momenta parallel to the $B$-field are distributed
according to Eq.~\eqref{eq:f_p}. The integration of the product of
this distribution with the corresponding scattering cross sections
leads to the cumulative distribution function (Eq.~\eqref{eq:F_p}).
During the MC simulation the scattering electron's parallel momentum
is drawn from this distribution. The influence of this sampling
process on the cyclotron line shape is illustrated in
Fig.~\ref{fig:pi}. It shows the momenta drawn during a simulation with
respect to the scattering photons' energies. The probability for
photons to scatter off an electron with the right momentum to make the
photon fulfill the resonance condition in the electron's rest frame is
much higher. Therefore the events shown in Fig.~\ref{fig:pi} track the
resonance condition \citep[see, e.g.,][]{daugherty78,harding91}
\begin{equation}\label{eq:p_zlw}
  \sqrt{ p^2c^2 + m_e^2c^4 } = pc \cos\vartheta + n \frac{B}{B_\mathrm{crit}} \frac{m_e^2c^4}{\omega} - \frac{\omega}{2} \sin^2\vartheta\,.
\end{equation}
\citet{harding91} already discussed the solutions to this equation,
which have to be found numerically and are often being referred to as
zero line width solutions. They are shown as curved lines in
Fig.~\ref{fig:pi}. The resonance condition can only be fulfilled for
photon energies below the cut-off energy,
\begin{equation}\label{eq:E_cut}
  \omega_\mathrm{cut}(n) =
  \frac{\sqrt{1+2nB/B_{\mathrm{crit}}}-1}{\sin\vartheta} m_e c^2\,.
\end{equation}
The vertical solid lines in Fig.~\ref{fig:pi} mark the corresponding
positions of the cut-off energies. No interaction involving the
excitation of an electron to Landau level $n$ occurs beyond that
energy. In the following we describe the zero line width solutions in
energy-momentum space and provide physically motivated regimes in
which these solutions are expected to be fulfilled, that is, energy
and momentum limits. We show, for the first time, simulated MC
scattering events in the context of these zero line width solutions
(and their limits) with the aim to establish an understanding of the
complex cyclotron scattering process. Furthermore, we use this
representation as a consistency check for our MC simulation.
\citet{schwarm_diploma} performed a similar check using the MC code by
\citet{araya97} without the physically motivated limits in energy and
momentum space. Only photons moving upwards or almost perpendicular to
the column axis are shown, which is why no decrease in electron
momentum occurs. Numerically calculated solutions of
Eq.~\eqref{eq:p_zlw} are shown in Fig.~\ref{fig:pi} as black solid
curves. Vertical solid lines mark the positions of the cut-off
energies corresponding to the three cyclotron resonances shown.
Positive and negative electron momenta are possible to fulfill the
resonance condition for photons with incoming angles almost
perpendicular to the magnetic field and energies below the cut-off
energy. At the cut-off energy a unique solution is possible for
electrons with zero parallel momentum. For smaller angles to the
magnetic field the momentum solutions deform, following the cut-off
energy which is shifting to higher energies. The sampled events are
bound to a limited region of the energy-momentum space. The cyclotron
resonance widths provide boundaries on the energy axes, since only
photons whose momentum averaged mean free path is small enough for
scattering to take place will appear in the figure. As an estimate, we
use the full Doppler width for thermal cyclotron line broadening
\citep{meszaros85a},
\begin{equation}\label{eq:E_width}
  \frac{\Delta E}{E} = \sqrt{ 8 \ln{2}\,\frac{k_\mathrm{B}T}{m_e c^2} } \cos\vartheta\,.
\end{equation}
The electron momentum space is limited by the width of the
relativistic Maxwellian momentum distribution. We use its definition
in Eq.~\eqref{eq:f_p} to calculate the momentum range in which $<
99$\% of the initial electrons reside. Therefore, the plasma
temperature is the main parameter responsible for a limited spreading
of scattering events in energy-momentum space \citep{meszaros78}. The
boundaries of the energy-momentum space available for scattering
events are shown in Fig.~\ref{fig:pi} as dotted vertical and
horizontal lines respectively.

Finally, the dashed lines in Fig.~\ref{fig:pi} show the cyclotron line
energies, that is, where $n(\omega,\mu)$ is an integer (see
Eq.~\eqref{eq:n_interp}). For photons moving perpendicular to the
$B$-field this line overlaps with the cut-off energy.

Comparing the red points in Fig.~\ref{fig:pi}, which mark the initial
parallel momentum, with the blue points, which correspond to the
electron momentum after the scattering process, shows that the
electrons gain parallel momentum in the (projected) direction of the
incoming photon. For angles perpendicular to the magnetic field there
is no continuous momentum transfer, while for photons incoming at
smaller angles to the magnetic field parallel momentum is transferred.
Figure~\ref{fig:pi} shows only photons moving upwards. The electron
momenta for scattering processes with a parallel component are
therefore shifted to higher positive momenta.

The number of scattering events declines for very small photon angles
to the magnetic field, from $\sim$4000 scattering events in the top
panel to $\sim$300 events in the bottom panel. The reason for this is
twofold: first, the almost zero photon momentum component
perpendicular to the field leads to a decrease of resonant cyclotron
scattering events and second, repeated cyclotron scattering favors a
redistribution of the interacting photons towards larger angles to the
B-field, especially for photon energies close to the cyclotron energy.
A detailed description of the classic CRSF geometries will be provided
in paper~II.

Now we are able to verify that scattering events are of resonant
nature (i.e., they involve an excitation of an electron), the resonant
scattering events coincide with the zero line width solutions, and the
final electron momentum is increasing with decreasing scattering
angle. The points which are not located on the zero line width
solutions mark elastic scattering processes not involving any electron
excitations. They occur mainly close to $\vartheta $$\sim$$ 90^\circ$
where resonant scattering is suppressed because the photons have to
have exactly the right energy to excite an electron. Photons with a
smaller angle to the magnetic field, on the other hand, can transfer
excessive energy to the electron's momentum parallel to the field.
Therefore the red and blue points of the resonant scattering events
are almost congruent in the top panel of Fig.~\ref{fig:pi}, while the
gap between them increases with increasing $\mu = \cos\vartheta$. The
data show that our physically motivated limits provide accurate
boundaries for the majority of scattering events and further justifies
the usage of Eq.~\ref{eq:E_width} as an approximation to the CRSF line
width.

\section{Summary}

For accelerating Monte Carlo simulations of cyclotron lines we
replaced the most time consuming part -- that is the calculation of
the photon mean free path -- with a tabular interpolation scheme.
These tables store the mean free paths of photons for different
incident angles and energies. The partial results necessary for
interpolating the electron momentum after a mean free path has been
drawn from the exponential probability distribution, and the spin
dependent results, are saved as well. The electronic tables described
here are available online.

This interpolation scheme is used to generate synthetic cyclotron line
spectra using our MC simulation code. It enables us to simulate much
more complex physical scenarios than previous works. As an example, we
have investigated the application of the momentum sampling.

\begin{acknowledgements}
  This work has been partially funded by the Deutsche
  Forschungsgemeinschaft under DFG grant number WI 1860/11-1 and by
  the Deutsches Zentrum f\"ur Luft- und Raumfahrt under DLR grant
  numbers 50\,OR\,1113, 50\,OR\,1207, and 50\,OR\,1411. MTW is supported
  by the Chief of Naval Research and by the National Aeronautics and Space 
  Administration Astrophysical Data Analysis Program. We thank the
  International Space Science Institute in Bern for inspiring team
  meetings. The fruitful discussions within the MAGNET collaboration
  also had a very positive impact on this work.
\end{acknowledgements}

\appendix

\section{Interpolation Techniques}\label{sec:interpo}
In the following we explain the interpolation techniques used for the
creation of the MFP tables and to interpolate from them during CRSF
simulations with our new MC code.

As discussed above, the interpolation tables were refined adaptively
by comparing the calculated profile $\sigma_\mathrm{calc}$ of a new
point with the profile obtained by interpolation from the table
$\sigma_\mathrm{interp}$, without taking this newly calculated point
into account. A linear interpolation scheme was used for refining the
energy grid, that is
\begin{equation}
  \label{interp_lin}
\langle\sigma\rangle_{\mathrm{interp}}(\mu, \omega) =
\langle\sigma\rangle(\mu, \omega_1) + \frac{\langle\sigma\rangle(\mu,
  \omega_2) - \langle\sigma\rangle(\mu, \omega_1)}{\omega_2 -
  \omega_1} (\omega - \omega_1)\,.
\end{equation}

The splitting of the energy intervals is stopped if
\begin{equation}
  \label{eq:interp_conv}
\frac{\langle\sigma\rangle_{\mathrm{interp}} - \langle\sigma\rangle_{\mathrm{calc}}}{\langle\sigma\rangle_{\mathrm{calc}}} \le \epsilon\,,
\end{equation}
where $\epsilon = 1/15$ has been found to be a good compromise between
precision and calculation time.

The same linear interpolation method can be used to interpolate
profiles during Monte Carlo simulations. For an interpolation in $\mu$
and $\omega$ two linear interpolations have to be performed. First the
inverse mean free path is interpolated for the surrounding angular
grid points, each one on its own energy grid. Linear interpolation
between these values with respect to the angle then gives the desired
value.

For using the computed mean free path tables for the simulation of
synthetic spectra with CRSFs, a physically motivated interpolation
method was implemented as well. Instead of interpolating for the same
energy on two different angular grid points, the energy is shifted to
the corresponding energy of the angular point considered. This is done
by calculating the order $n$ of the energy in terms of the resonance
condition for the desired angular value. This value is used to
calculate the corresponding energies at the angular boundaries used
for interpolation.
\begin{equation}
  \label{eq:n_interp}
n(\omega,\mu) = \frac{{(\omega \sin^2 \vartheta / m_e c^2 + 1)}^2 - 1} {2 B/B_\mathrm{crit} \sin^2 \vartheta}
\end{equation}
This interpolation scheme eliminates inaccuracies due to the shift of
the resonance energy with $\vartheta$, because interpolation is done
at nearly constant ``elevation'' along the resonance ridges
\citep{harding91}.

\section{Mean free path table structure}\label{Appendix:mfptables}

The naming convention of the mean free path tables was chosen as
follows: All names start with \texttt{mfp\_} followed by the table
specific parameters, namely the magnetic field strength $B$ and the
temperature $T$. The magnetic field is in units of the critical
magnetic field strength, with five digits The temperature
$k_\mathrm{B}T$ is given in units of MeV with five digits as well. For
example, the first table in alphabetical order is named
\texttt{mfp\_B0.0100T0.0030.fits}.

Each mean free path table starts with an empty image HDU containing
header keywords for the magnetic field strength $B$, the temperature
$T$, and the relative maximum error as listed in
table~\ref{table:mfp_keys}. It is followed by a variable number of
binary extensions any of which corresponds to one angular grid point.
These binary extensions are ordered by increasing $\mu =
\cos\vartheta$ and are described in detail in the following.

The value of $\mu$ is stored in the header keyword \texttt{MU}. Each
row corresponds to one energy grid point, the value of which can be
found in the first column. The second column contains the total
thermally averaged scattering cross section summed over final electron
spin states. The number of grid points used for the adaptive Simpson
integration over the thermal electron momentum distribution
(Eq.~\eqref{eq:profile}), can be found in the third column. The
corresponding electron momentum grid points are stored in the fourth
column as a variable length array of double numbers, followed by the
spin averaged cumulative distribution function (CDF), the ensemble of
partial integrals obtained from Eq.~\eqref{eq:F_p}. The sixth column
contains, again, the number of electron momentum grid points used for
the integration of Eq.~\eqref{eq:profile} taking into account only
transitions with final electron spin down. Following the corresponding
momentum grid points and cumulative distribution functions the pattern
repeats once again for the case of final electron spin up.

The header keyword \texttt{MAX\_ERR} contains the relative error of
the table, that is, $\epsilon$ from Eq.~\ref{eq:interp_conv}, in units
of $1/15$.

\begin{table}
  \caption{List of FITS keywords.}\label{table:mfp_keys}
  \centering
  \begin{tabular}{cp{0.7\columnwidth}}
\hline\hline
Name & Description \\
\hline
B & Magnetic field [$B_{\mathrm{crit}}$] \\
T & Electron temperature $k_\mathrm{B}T$ [MeV] \\
MU & Cosine of the incoming photons angle to the magnetic field, $\mu = \cos\vartheta$ \\
MAX\_ERR & Maximum relative error in units of 1/15 \\
\hline
  \end{tabular}
\end{table}

Table~\ref{table:mfp_struct} visualizes the structure described here.
The last six columns for the spin dependent cases are omitted for the
sake of clarity.

\begin{table}
  \caption{Uncompressed file sizes for the parameter combinations made
    available in GB. Compression reduces the file size by
    approximately 50\%.}\label{table:mfp_list} \centering
  \begin{tabular}{cccccc}
\hline
 & \multicolumn{5}{c}{$B/B_\mathrm{crit}$} \\
$k_\mathrm{B}T$ [keV]     & 0.01 & 0.03 & 0.06 & 0.09 & 0.12 \\
\hline\hline
$\phantom{0}3$ & $\phantom{0}44$  &  16  & 10   & $\phantom{0}7$   &  11 \\
$\phantom{0}6$ & $\phantom{0}89$  &  29  & 16   & 11   &  16 \\
$\phantom{0}9$ & 148  &  42  & 23   & 16   &  22 \\ 
12             & 198  &  61  & 32   & 20   &  26 \\
15             & 238  &  74  & 35   & 24   &  32 \\
\hline
\end{tabular}
\end{table}

\begin{table*}
  \caption{Description of the mean free path table structure for given
    values of $k_\mathrm{B}T$ and $B / B_\mathrm{crit}$. The column
    pattern \texttt{\# of grid points}, \texttt{Grid [MeV]},
    \texttt{CDF} is repeated two more times, omitted below for
    clarity, for the case of final electron spin down and final
    electron spin up, respectively. The first element of each
    \texttt{Grid} and \texttt{CDF} array is used for internal
    consistency checks. The user should only use the indices 1 to
    \texttt{$N_p$}. The dependency of \texttt{$N_p$} on angle and
    energy is not stated explicitly. $N_\mu$, $N_k$, and $N_p$ are
    typically on the order of a few hundred to thousand of points.
  }\label{table:mfp_struct} \centering
  \begin{tabular}{c c c c c c c}
\hline\hline
EXTENSION & MU & & & COLUMNS & & \\
\hline
 & & Energy [MeV] & Cross section [$\sigma_{\mathrm{Th}}$] & \# of grid points & Grid [MeV] & CDF \\
\hline
1         & $\mu_1$ & $k_1$ & $\langle\sigma(\mu_1,k_1)\rangle$ & $N_p$ & $[p_1, \dotsc ,p_{N_p}]$ & $[F_{-m_ec}^{p_1}, \dotsc , F_{-m_ec}^{p_{N_p}}]$ \\
          &         & $\vdots$ & & & & \\
          & & $k_{N_k}$ & $\langle\sigma(\mu_1,k_{N_k})\rangle$ & $N_p$ & $[p_1, \dotsc ,p_{N_p}]$ & $[F_{-m_ec}^{p_1}, \dotsc , F_{-m_ec}^{p_{N_p}}]$ \\
$\vdots$  & $\vdots$ & $\vdots\vdots$ & $\vdots\vdots$ & $\vdots\vdots$ & $\vdots\vdots$ & $\vdots\vdots$\\
$N_{\mu}$ & $\mu_{N_{\mu}}$ & $k_1$ & $\langle\sigma(\mu_{N_{\mu}},k_1)\rangle$ & $N_p$ & $[p_1, \dotsc ,p_{N_p}]$ & $[F_{-m_ec}^{p_1}, \dotsc , F_{-m_ec}^{p_{N_p}}]$ \\
          &         & $\vdots$ & & & & \\
          & & $k_{N_k}$ & $\langle\sigma(\mu_{N_{\mu}},k_{N_k})\rangle$ & $N_p$ & $[p_1, \dotsc ,p_{N_p}]$ & $[F_{-m_ec}^{p_1}, \dotsc , F_{-m_ec}^{p_{N_p}}]$ \\
\hline
  \end{tabular}
\end{table*}

\begin{figure*}
  \includegraphics[width=17cm]{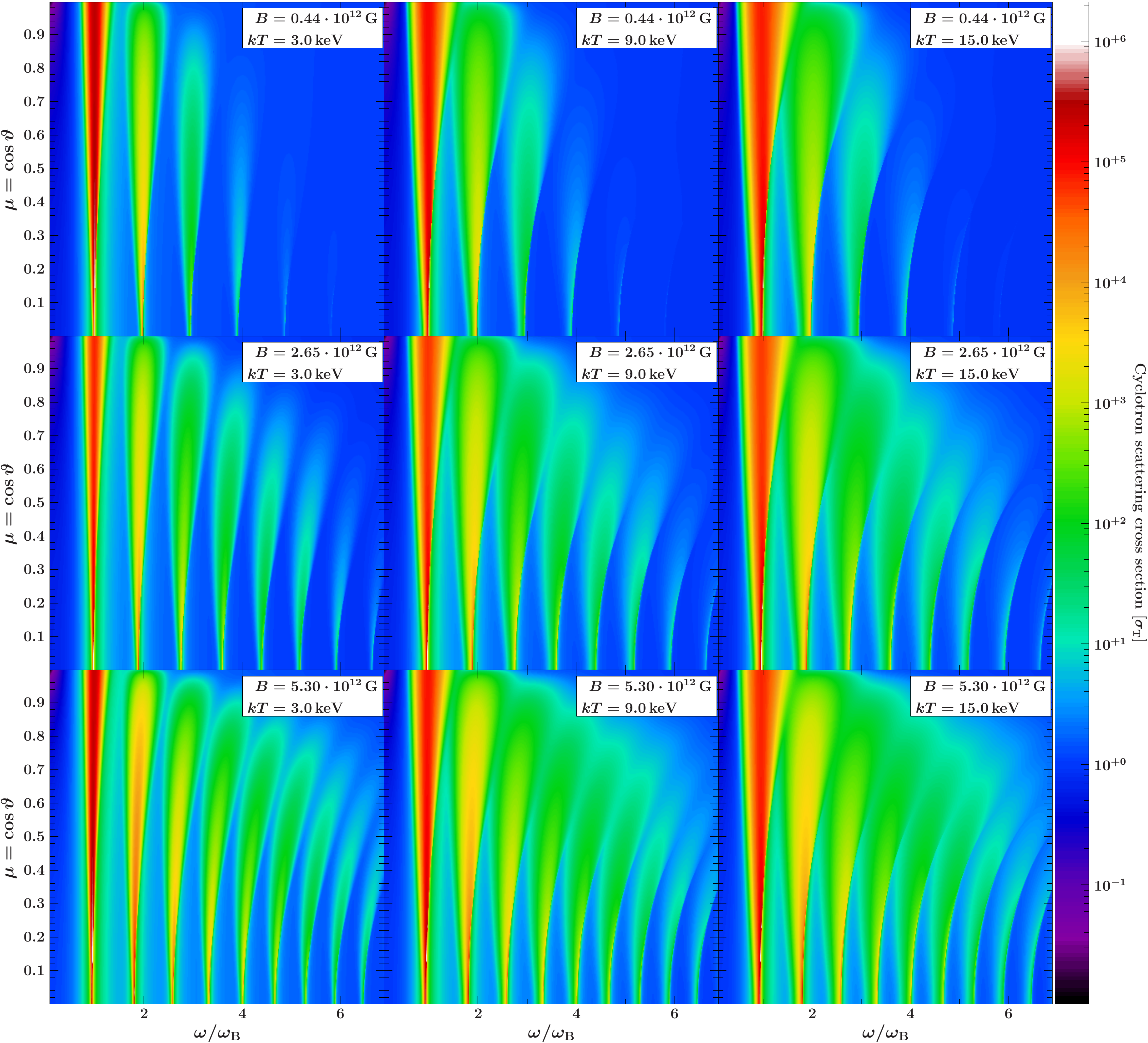}
  \caption{ Visualization of the adaptively calculated inverse mean
    free path values from the interpolation tables for $B=0.01$, 0.06,
    and $0.12\,B_\mathrm{crit}$. The colors correspond to the averaged
    scattering cross sections. }\label{fig:mfp_tables}
\end{figure*}

\begin{figure*}
\centering
  \includegraphics[width=17cm]{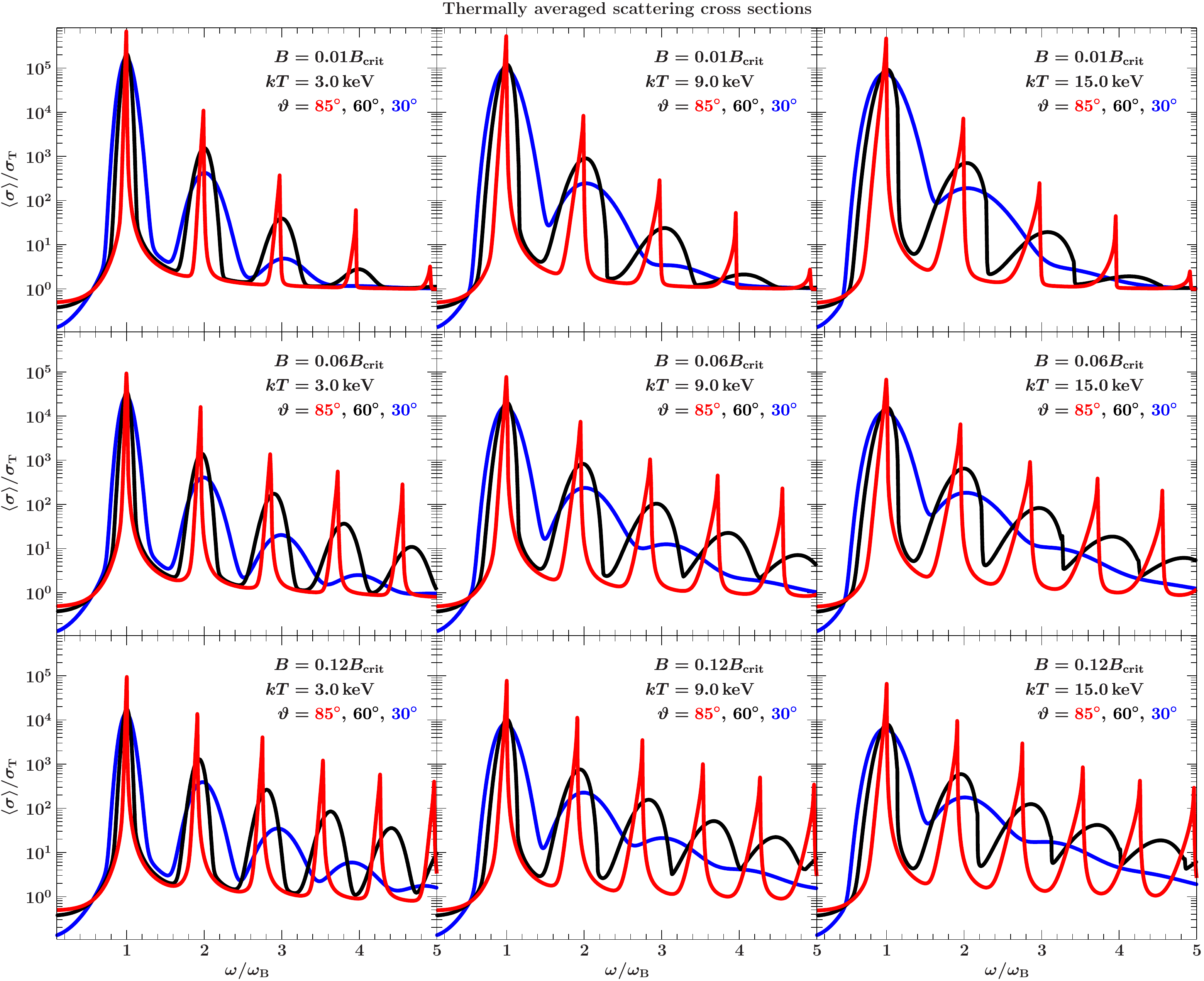}
  \caption{ Adaptively calculated averaged cyclotron resonance
    scattering cross sections. The colors correspond to different
    angles of the scattering photon with respect to the magnetic
    field. }\label{fig:mfp_profiles}
\end{figure*}


\begin{thebibliography}{28}
\expandafter\ifx\csname natexlab\endcsname\relax\def\natexlab#1{#1}\fi

\bibitem[{{Araya}(1997)}]{araya97}
{Araya}, R.~A. 1997, PhD thesis, Johns Hopkins University, Baltimore

\bibitem[{{Araya} \& {Harding}(1999)}]{araya99}
{Araya}, R.~A. \& {Harding}, A.~K. 1999, \apj, 517, 334

\bibitem[{{Becker} \& {Wolff}(2007)}]{becker07}
{Becker}, P.~A. \& {Wolff}, M.~T. 2007, \apj, 654, 435

\bibitem[{{Bonazzola} {et~al.}(1979){Bonazzola}, {Heyvaerts}, \&
  {Puget}}]{bonazzola79}
{Bonazzola}, S., {Heyvaerts}, J., \& {Puget}, J.~L. 1979, \aap, 78, 53

\bibitem[{{Bussard} {et~al.}(1986){Bussard}, {Alexander}, \&
  {Meszaros}}]{bussard86}
{Bussard}, R.~W., {Alexander}, S.~B., \& {Meszaros}, P. 1986, \prd, 34, 440

\bibitem[{{Canuto} {et~al.}(1971){Canuto}, {Lodenquai}, \&
  {Ruderman}}]{canuto71}
{Canuto}, V., {Lodenquai}, J., \& {Ruderman}, M. 1971, \prd, 3, 2303

\bibitem[{{Daugherty} \& {Harding}(1986)}]{daugherty86}
{Daugherty}, J.~K. \& {Harding}, A.~K. 1986, \apj, 309, 362

\bibitem[{{Daugherty} \& {Ventura}(1978)}]{daugherty78}
{Daugherty}, J.~K. \& {Ventura}, J. 1978, \prd, 18, 1053

\bibitem[{{Gonthier} {et~al.}(2014){Gonthier}, {Baring}, {Eiles}, {Wadiasingh},
  {Taylor}, \& {Fitch}}]{gonthier14}
{Gonthier}, P.~L., {Baring}, M.~G., {Eiles}, M.~T., {et~al.} 2014, \prd, 90,
  043014

\bibitem[{{Graziani}(1993)}]{graziani93}
{Graziani}, C. 1993, \apj, 412, 351

\bibitem[{{Harding} \& {Daugherty}(1991)}]{harding91}
{Harding}, A.~K. \& {Daugherty}, J.~K. 1991, \apj, 374, 687

\bibitem[{{Herold} {et~al.}(1982){Herold}, {Ruder}, \& {Wunner}}]{herold82}
{Herold}, H., {Ruder}, H., \& {Wunner}, G. 1982, \aap, 115, 90

\bibitem[{{Isenberg} {et~al.}(1998){Isenberg}, {Lamb}, \& {Wang}}]{isenberg98}
{Isenberg}, M., {Lamb}, D.~Q., \& {Wang}, J.~C.~L. 1998, \apj, 505, 688

\bibitem[{{Johnson} \& {Lippmann}(1949)}]{johnson49}
{Johnson}, M.~H. \& {Lippmann}, B.~A. 1949, Phys.\ Rev., 76, 828

\bibitem[{{Langer}(1981)}]{langer81}
{Langer}, S.~H. 1981, \prd, 23, 328

\bibitem[{{Latal}(1986)}]{latal86}
{Latal}, H.~G. 1986, \apj, 309, 372

\bibitem[{Lyness(1969)}]{lyness69}
Lyness, J.~N. 1969, J. ACM, 16, 483

\bibitem[{McKeeman(1962)}]{mckeeman62}
McKeeman, W.~M. 1962, Comm.\ ACM, 5, 604

\bibitem[{{M\'esz\'aros}(1978)}]{meszaros78}
{M\'esz\'aros}, P. 1978, \aap, 63, L19

\bibitem[{{Meszaros} \& {Nagel}(1985)}]{meszaros85a}
{Meszaros}, P. \& {Nagel}, W. 1985, \apj, 298, 147

\bibitem[{{M{\'e}sz{\'a}ros} \& {Ventura}(1978)}]{meszaros78a}
{M{\'e}sz{\'a}ros}, P. \& {Ventura}, J. 1978, Physical Review Letters, 41

\bibitem[{{MPI Forum}(1994)}]{mpi94}
{MPI Forum}. 1994, MPI: A Message-Passing Interface Standard, {Technical Report
  FUT-CS-94-230}, University of Tennessee, Knoxville, TN

\bibitem[{{Sch{\"o}nherr} {et~al.}(2007){Sch{\"o}nherr}, {Wilms}, {Kretschmar},
  {Kreykenbohm}, {Santangelo}, {Rothschild}, {Coburn}, \&
  {Staubert}}]{schoenherr07}
{Sch{\"o}nherr}, G., {Wilms}, J., {Kretschmar}, P., {et~al.} 2007, \aap, 472,
  353

\bibitem[{{Schwarm} {et~al.}(2012){Schwarm}, {Sch{\"o}nherr}, {Wilms}, \&
  {Kretschmar}}]{schwarm12}
{Schwarm}, F., {Sch{\"o}nherr}, G., {Wilms}, J., \& {Kretschmar}, P. 2012, PoS,
  INTEGRAL 2012, 153

\bibitem[{{Schwarm}(2010)}]{schwarm_diploma}
{Schwarm}, F.-W. 2010, Diploma thesis, Univ.~Erlangen-Nuremberg

\bibitem[{{Schwarm et al.}(2016)}]{schwarm16b}
{Schwarm et al.} 2016, \aap, to be submitted (paper II)

\bibitem[{{Sina}(1996)}]{sina96}
{Sina}, R. 1996, PhD thesis, University of Maryland, College Park, MD

\bibitem[{{Sokolov} {et~al.}(1968){Sokolov}, {Ternov}, {Bagrov}, {Gal'tsov}, \&
  {Zhukovskii}}]{sokolov68}
{Sokolov}, A.~A., {Ternov}, I.~M., {Bagrov}, V.~G., {Gal'tsov}, D.~V., \&
  {Zhukovskii}, V.~C. 1968, Soviet Physics, 11, 4

\end{thebibliography}
\end{document}